\documentclass[graybox]{svmult}

\smartqed
\usepackage{mathptmx}       %
\usepackage{helvet}         %
\usepackage{courier}        %
\usepackage{type1cm}        %
\usepackage{graphicx}       %
\graphicspath{{Images/}}

\usepackage{array,colortbl}
\usepackage{array}
\usepackage{amsmath,amsfonts,amssymb,bm} %
\DeclareFontFamily{U}{mathx}{\hyphenchar\font45}
\DeclareFontShape{U}{mathx}{m}{n}{
      <5> <6> <7> <8> <9> <10>
      <10.95> <12> <14.4> <17.28> <20.74> <24.88>
      mathx10
      }{}
\DeclareSymbolFont{mathx}{U}{mathx}{m}{n}
\DeclareFontSubstitution{U}{mathx}{m}{n}
\DeclareMathAccent{\widecheck}      {0}{mathx}{"71}

\renewcommand{\email}[1]{\emailname: #1} % change the email address font style

\usepackage{mathptmx}       % selects Times Roman as basic font
\usepackage{helvet}         % selects Helvetica as sans-serif font
\usepackage{courier}        % selects Courier as typewriter font
\usepackage{makeidx}         % allows index generation
\usepackage{graphicx}        % standard LaTeX graphics tool
\usepackage[bottom]{footmisc}% places footnotes at page bottom

\usepackage{latexsym}
\usepackage{amsmath}
\usepackage{amsfonts}
\usepackage{amssymb}
\usepackage{bm}

\usepackage{url}
\usepackage{algorithm}
\usepackage{algorithmic}
\usepackage[misc,geometry]{ifsym}

\spdefaulttheorem{assumption}{Assumption}{\upshape \bfseries}{\itshape}
\spdefaulttheorem{algo}{Algorithm}{\upshape \bfseries}{\itshape}
\DeclareSymbolFont{bbold}{U}{bbold}{m}{n}
\DeclareSymbolFontAlphabet{\mathbbold}{bbold}

\DeclareSymbolFont{bbold}{U}{bbold}{m}{n}
\DeclareSymbolFontAlphabet{\mathbbold}{bbold}

\usepackage{microtype} %

\usepackage[colorlinks=true,linkcolor=black,citecolor=black,urlcolor=black]{hyperref}
\usepackage{bookmark}
\makeatletter %
  \providecommand*{\toclevel@author}{999}
  \providecommand*{\toclevel@title}{0}
\makeatother

\usepackage{caption}
\usepackage{subcaption}
\usepackage{tikz}
\usetikzlibrary{arrows,shapes,positioning}
\usepackage{pgfplots}
\usepgfplotslibrary{fillbetween}
\usepackage{cleveref} %
\crefname{lstlisting}{listing}{listings}
\usepackage{listings}
\usepackage{comment}
\excludecomment{minicg}
\begin{minicg}
	output_into_include_files = True
	from tikz import Tikz
	from fragmented_radixtree import RadixTree
	import random

	random.seed(42)
	colors = ["HighlightColor", "HighlightColor2", "HighlightColor3", "HighlightColor4", "GoogleColor", "HighlightColor5", "AmazonColor", "AMDColor", "IntelColor", "AppleColor"]
\end{minicg}

\definecolor{HighlightColor}{rgb}{0.462745098, 0.725490196, 0.000000000}
\definecolor{HighlightColor2}{rgb}{0.000000000, 0.584313725, 0.466666667}
\definecolor{HighlightColor3}{rgb}{0.447058824, 0.168627451, 0.580392157}
\definecolor{HighlightColor4}{rgb}{0.043137255, 0.298039216, 0.788235294}
\definecolor{HighlightColor5}{rgb}{0.360784314, 0.360784314, 0.360784314}
\definecolor{IntelColor}{rgb}{0.000000000, 0.525490196, 0.815686275}
\definecolor{AMDColor}{rgb}{0.505882353, 0.505882353, 0.505882353}
\definecolor{AppleColor}{rgb}{0.752941176, 0.752941176, 0.752941176}
\definecolor{GoogleColor}{rgb}{0.894117647, 0.28627451, 0.270588235}
\definecolor{AmazonColor}{rgb}{1.000000000, 0.662745098, 0.000000000}
\definecolor{DarkGray}{rgb}{0.360784314, 0.360784314, 0.360784314}
\definecolor{MediumGray}{rgb}{0.505882353, 0.505882353, 0.505882353}
\definecolor{LightGray}{rgb}{0.752941176, 0.752941176, 0.752941176}
\definecolor{VeryLightGray}{rgb}{0.905882353, 0.905882353, 0.905882353}

\begin{document}

\title*{Massively Parallel Construction of Radix Tree Forests for the Efficient Sampling of Discrete or Piecewise Constant
Probability Distributions}
\titlerunning{Radix Tree Forests for the Efficient Sampling
Probability Distributions
}
\author{Nikolaus Binder \and Alexander Keller}
\institute{
Nikolaus Binder \and Alexander Keller
\at NVIDIA
\email{nbinder@nvidia.com}, \email{akeller@nvidia.com}
}
\maketitle

\abstract{We compare different methods for sampling from discrete or piecewise constant probability distributions and introduce a new algorithm
which is especially efficient on massively parallel processors, such as GPUs. The scheme preserves the distribution
properties of the input sequence, exposes constant time complexity on the average,
and significantly lowers the average number of operations for certain distributions when sampling is performed in a
parallel algorithm that requires synchronization.
Avoiding load balancing issues of na\"ive approaches,
a very efficient massively parallel construction algorithm
for the required auxiliary data structure is proposed.}

\section{Introduction}
\label{Sec:introduction}

\begin{figure}[h]
	\includegraphics[width=\linewidth]{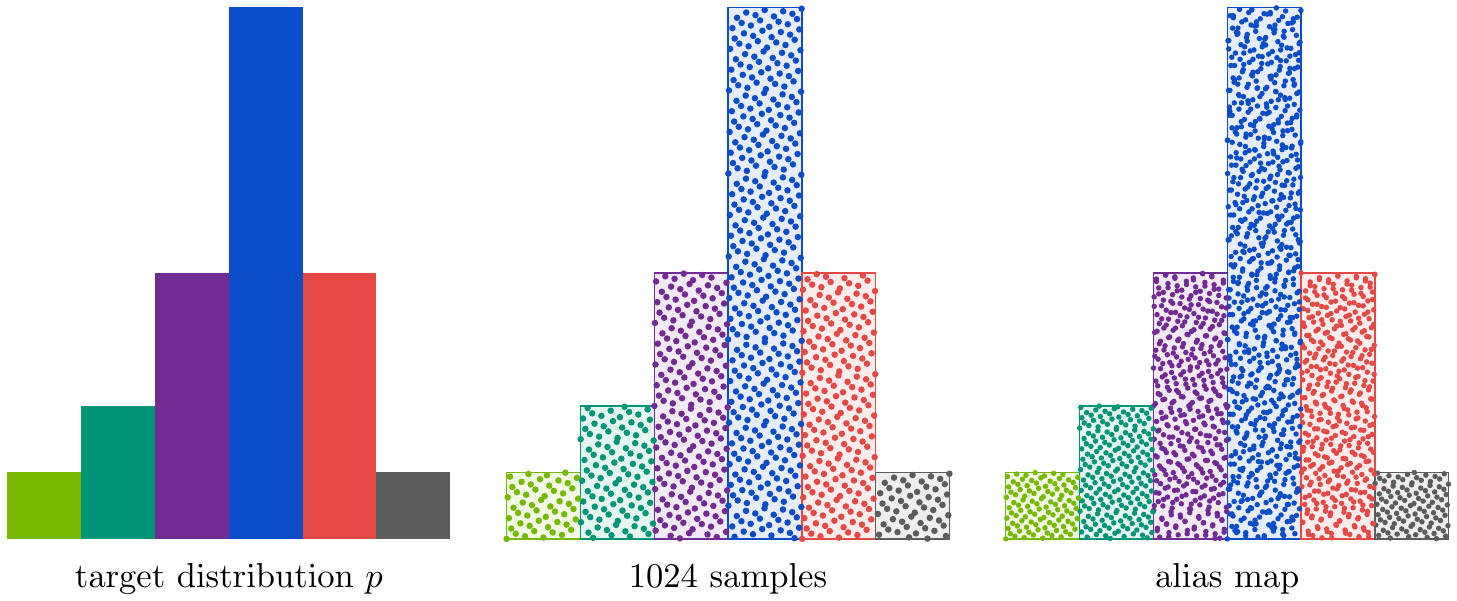}
	\caption{Drawing samples according to a given
	piecewise constant probability distribution (left) using a monotonic mapping can preserve
	uniformity properties, for example a discrepancy, %
	of an input sample sequence (middle), while using a non-monotonic mapping such as the Alias Method \cite{Walker1974,Walker1977}
	negatively affects uniformity (right). For illustrative purposes we distribute the samples in columns with heights proportional to the
	probabilities. Then, the point set just is partitioned into the columns and scaling the
	columns to equal heights would yield the desired density distribution. The input for both
	mappings is the two-dimensional Hammersley set with 1024 points. The first component is mapped to
	the index of the column, using the fractional part
	for the $y$ position. For illustrative purposes, we use the second component to
	distribute the points also along the $x$ axis inside each column. It is important to note that samples
	in higher columns are more often mapped to an alias and
	therefore become less uniform.
	}
	\label{Fig:Teaser}
\end{figure}

In many applications, samples need to be drawn according to a given discrete probability density
\[
  p := (p_1, p_2, \ldots, p_n) ,
\]
where the $p_i$ are positive and $\sum_{i = 1}^n p_i  = 1$.
Defining $P_0 :=0$ and the partial sums $P_k := \sum_{i = 1}^k p_i$ results in
$
  0 = P_0 < P_1 < \cdots < P_n = 1
$,
which forms a partition of the unit interval $[0,1)$.
Then the inverse cumulative distribution function (CDF)
\[
  P^{-1}(x) = i \Leftrightarrow P_{i - 1} \leq x < P_i
\]
can be used to map realizations of a uniform random variable $\xi$ on $[0, 1)$
to $\{1, 2, \ldots, n\}$ such that
\[
  \text{Prob}\left(\{ P_{i -1} \leq \xi < P_i\}\right) = p_i .
\]

Besides identifying the most efficient method to perform such a mapping, we are
interested in transforming low discrepancy
sequences \cite{Nie:92} and how such mappings affect
the uniformity of the resulting warped sequence. %
An example for such a sampling process is shown in \Cref{Fig:Teaser}.

The remainder of the article is organized as follows:
After reviewing several algorithms to sample according to
a given discrete probability density
by transforming uniformly distributed samples in \Cref{Sec:Previous_work},
massively parallel algorithms to construct auxiliary data structures for the
accelerated computation of $P^{-1}$ are introduced in \Cref{Sec:Method}.
The results of the scheme that preserves distribution properties, especially
when transforming low discrepancy sequences, are presented in \Cref{Sec:Results}
and discussed in \Cref{Sec:Discussion} before drawing the conclusions.

\section{Sampling from Discrete Probability Densities}
\label{Sec:Previous_work}

In the following we will survey existing methods to evaluate the inverse mapping $P^{-1}$ and compare their
properties with respect to computational complexity, memory requirements, memory access patterns,
and sampling efficiency.

\subsection{Linear Search}
\label{Sec:linear_search}

\begin{figure}[h]
	\includegraphics[width=\linewidth]{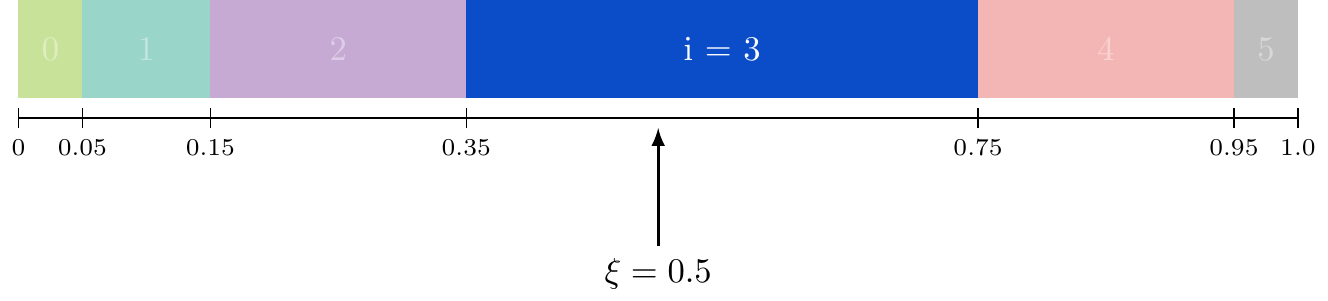}
	\caption{Linearly searching through all intervals until the interval containing the variable $\xi$ is found,
	requires ${\mathcal O}(n)$ steps on the average. In this example four comparison operations are required
to identify the interval $i=3$ that includes $\xi$.}
	\label{Fig:linear_search}
\end{figure}

As illustrated by the example in \Cref{Fig:linear_search}, a linear search computes the inverse
mapping $P^{-1}$ by subsequently checking all intervals for inclusion
of the value of the uniformly distributed
variable $\xi$. This is simple, does not require additional memory, achieves very good performance for
a small number $n$ of values, and scans the memory in linear order.
However, its average and worst case complexity of ${\mathcal O}(n)$ makes it
unsuitable for large $n$.

\subsection{Binary Search}
\label{Sec:binary_search}

\begin{figure}
	\includegraphics[width=\linewidth]{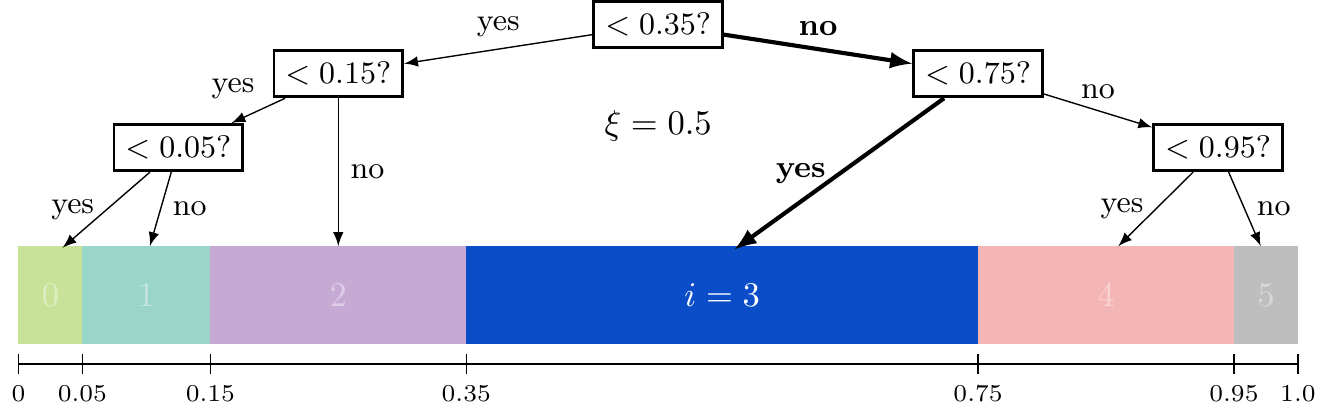}
	\caption{Binary search bisects the list of potential intervals in each step and hence has an average case and worst case
	complexity of ${\mathcal O}(\log_2 n)$. In the example two comparisons are required to find
	that $\xi = 0.5$ is included in the interval $i = 3$.}
	\label{Fig:binary_search}
\end{figure}

Binary search lowers the average and worst case complexity of the inverse mapping $P^{-1}$ to ${\mathcal O}(\log_2 n)$ by
performing bisection. Again, no additional memory is required, but memory no longer is accessed in linear order.
\Cref{Fig:binary_search} shows an example of binary search.

\subsection{Binary Trees}
\label{Sec:binary_trees}

The implicit decision tree traversed by binary search can be stored as an explicit binary tree
structure. The leaves of the tree reference intervals, and each node stores the value $q_i$
as well as references to the two children. The tree is then traversed by comparing $\xi$ to $q_i$ in each node,
advancing to the left child if $\xi < q_i$, and otherwise to the right child. Note that a clever enumeration scheme
could use the index $i$ for the node that splits the unit interval in $q_i$, and hence avoid explicitly storing $q_i$ in
the node data structure.

While the average case and worst case complexity of performing the inverse mapping $P^{-1}$ with such an explicitly stored
tree does not get below ${\mathcal O}(\log_2 n)$, allowing for arbitrary tree structures enables further optimization.
Again, memory access is not in linear order, but due to the information
required to identify the two children of each node in the tree, ${\mathcal O}(n)$ additional memory must be allocated and
transferred. It is important to note that the worst case complexity may be as high as ${\mathcal O}(n)$ for degenerate trees.
Such cases can be identified and avoided by an adapted re-generation of the tree.

\subsection{$k$-ary Trees}
\label{Sec:n_ary_trees}

On most hardware architectures, the smallest granularity
of memory transfer is almost always larger than what would be needed to perform a single comparison and to load the index of
one or the other child in a binary tree. 
The average case complexity for a branching factor $k$ is ${\mathcal O}(log_k n)$, but the
number of comparisons is either increased to $log_2 k$ (binary search in children) or even $k - 1$ (linear scan over the
children) in each step, and therefore either equal or greater than for binary trees on the average.
Even though more comparisons are performed on the average,
it may be beneficial to use trees with a branching factor higher than two, because the
additional effort in computation may be negligible as compared to the memory transfer
that happens anyhow.

\begin{figure}[h]
	\includegraphics[width=\linewidth]{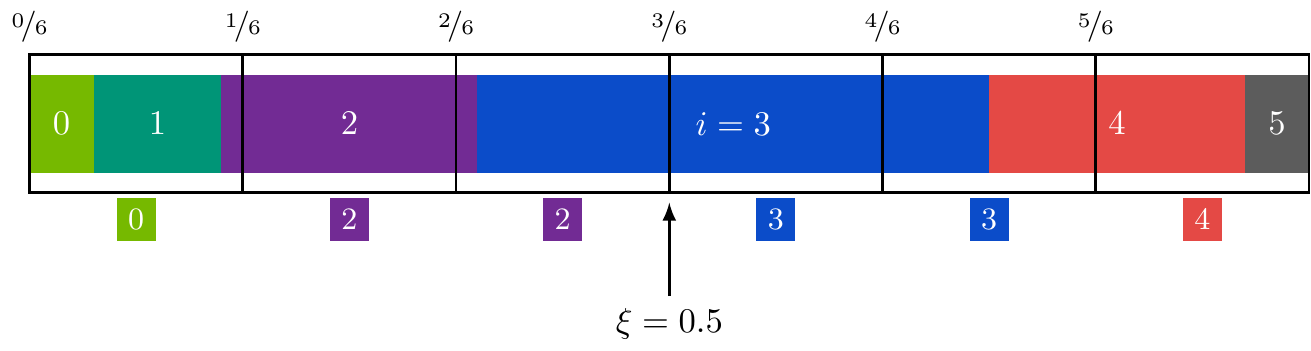}
	\caption{The Cutpoint Method uses a guide table that uniformly partitions the unit interval and stores the first
	interval that overlaps each cell (shown below each cell). These indices are then used as starting points for linear
	search. In this example only a single lookup is required; in general the average case complexity is ${\mathcal O}(1)$.}
	\label{Fig:cutpoint_method}
\end{figure}

\subsection{The Cutpoint Method}
\label{Sec:cutpoint_method}

By employing additional memory, the indexed search \cite{ChenAsau1974}, also known as the
Cutpoint Method \cite{Fishman1984}, can perform the inverse mapping $P^{-1}$ in
${\mathcal O}(1)$ on the average. Therefore, the unit interval is partitioned into $m$ cells of equal size and a \emph{guide
table} stores each first interval that overlaps a cell as illustrated in \Cref{Fig:cutpoint_method}. Starting with this
interval, linear search is employed to find the one that includes the realization of $\xi$.
As shown in \cite[Chapter 2.4]{Devroye:86}, the expected number of comparisons is
$1 + \frac{n}{m}$.
In the worst case all but one interval are
located in a single cell and since linear search has a complexity of ${\mathcal O}(n)$, the worst case complexity of the Cutpoint Method
is also ${\mathcal O}(n)$. Sometimes, these worst cases can be avoided by recursively nesting another guide table in cells with many entries.
In general, however, the problem persists. If nesting is performed multiple times, the structure of the nested guide
tables is similar to a $k$-ary tree (see \Cref{Sec:n_ary_trees}) with implicit split values defined by the equidistant partitioning.

Another way of improving the worst case performance is using binary search instead of linear search in each cell of the
guide table. No additional data needs to be stored since the index of the first interval of the next cell can be
conservatively used as the last interval of the current cell. The resulting complexity remains ${\mathcal O}(1)$ on the average,
but improves to ${\mathcal O}(\log_2 n)$ in the worst case.

\begin{figure}[h]
	\includegraphics[width=\linewidth]{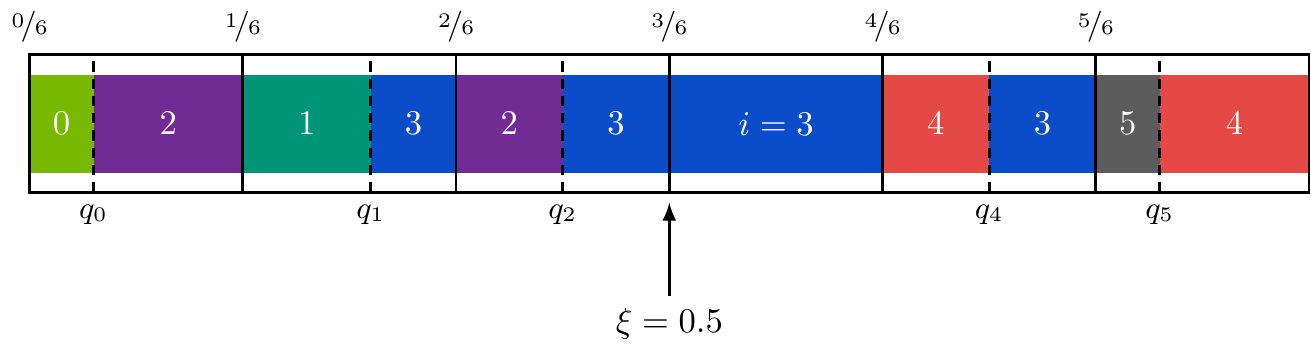}
	\caption{Similar to the Cutpoint Method, the Alias Method uses an additional table that uniformly partitions the unit
	interval. However, the Alias Method cuts and redistributes the intervals so that each cell contains at most two
	intervals, the first one being the interval with the same index as the cell and a second one that covers the rest of
	the range of the cell. Then, by storing the index of the second interval in each cell and the split points $q_j$ %
	of each cell with index $j$, the mapping can \emph{always} be evaluated in ${\mathcal O}(1)$, and in our example, as before, only a single lookup is
	required to find the interval $i = 3$ including $\xi = 0.5$.} 
	\label{Fig:alias_map}
\end{figure}

\subsection{The Alias Method}
\label{Sec:alias_map}

Using the Alias Method \cite{Walker1974,Walker1977}, the inverse $P^{-1}$ can be sampled in ${\mathcal O}(1)$ both on the average as well as
in the worst case. It avoids the worst case by cutting and reordering the intervals such that each cell contains at
most two intervals, and thus neither linear search nor binary search is required. For the example used in this article, a resulting table is
shown in \Cref{Fig:alias_map}. In terms of run time as well as efficiency of memory access the method is very compelling
since the mapping can be performed using a single read operation of exactly two values and one comparison, whereas
hierarchical structures generally suffer from issues caused by scattered read operations.

\begin{figure}[h]
	\centering
	\includegraphics[width = .75\linewidth]{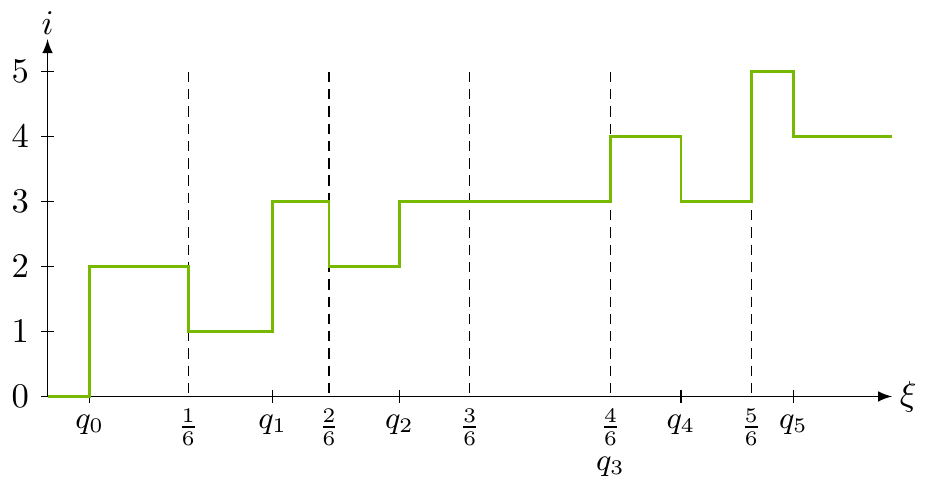}
	\caption{The mapping of the Alias Method is not the inverse of $P$. While it still maps a uniform variate to one with
the desired density distribution, it is  not monotonic. This instance is a possible mapping $\xi \mapsto
i$ for the Alias Method and the distribution shown in \Cref{Fig:Teaser}, where $i = m$ for $\xi \in \left[\frac{m}{M} , \frac{m
+ 1}{M}\right)$ for $\xi < q_m$, and $i = \textrm{alias}(m)$ for $\xi \geq q_m$. Since $\textrm{alias}(m) \neq m + 1$ in general, the mapping cannot be monotonic.
}
	\label{Fig:alias_map_details}
\end{figure}

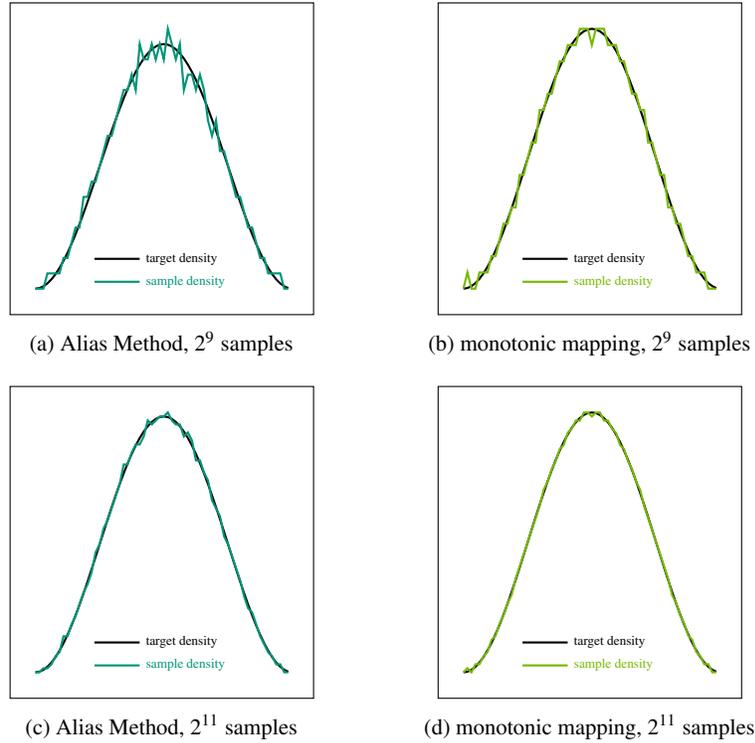
\begin{figure}[h]
	\begin{subfigure}{0.48\textwidth}
		\centering
		\begin{tikzpicture}
			\begin{axis}[
				ticks=none,
				legend style={draw=none},
				legend style={at={(0.5,0.05)},anchor=south, fill=none, font=\tiny},
				legend cell align=left,
				width=\textwidth,
				height=.3\textheight]
				\addlegendimage{empty legend}
				\addlegendentry{}
				\addplot[color=black, name path=f, thick]%
				coordinates {
					(0, 0.000000)
					(1, 0.000075)
					(2, 0.000300)
					(3, 0.000673)
					(4, 0.001189)
					(5, 0.001845)
					(6, 0.002633)
					(7, 0.003547)
					(8, 0.004576)
					(9, 0.005713)
					(10, 0.006944)
					(11, 0.008259)
					(12, 0.009646)
					(13, 0.011089)
					(14, 0.012577)
					(15, 0.014093)
					(16, 0.015625)
					(17, 0.017157)
					(18, 0.018673)
					(19, 0.020161)
					(20, 0.021604)
					(21, 0.022991)
					(22, 0.024306)
					(23, 0.025537)
					(24, 0.026674)
					(25, 0.027703)
					(26, 0.028617)
					(27, 0.029405)
					(28, 0.030061)
					(29, 0.030577)
					(30, 0.030950)
					(31, 0.031175)
					(32, 0.031250)
					(33, 0.031175)
					(34, 0.030950)
					(35, 0.030577)
					(36, 0.030061)
					(37, 0.029405)
					(38, 0.028617)
					(39, 0.027703)
					(40, 0.026674)
					(41, 0.025537)
					(42, 0.024306)
					(43, 0.022991)
					(44, 0.021604)
					(45, 0.020161)
					(46, 0.018673)
					(47, 0.017157)
					(48, 0.015625)
					(49, 0.014093)
					(50, 0.012577)
					(51, 0.011089)
					(52, 0.009646)
					(53, 0.008259)
					(54, 0.006944)
					(55, 0.005713)
					(56, 0.004576)
					(57, 0.003547)
					(58, 0.002633)
					(59, 0.001845)
					(60, 0.001189)
					(61, 0.000673)
					(62, 0.000300)
					(63, 0.000075)
				};%
				\addlegendentry{\textcolor{black}{target density}}

				\addplot[color=HighlightColor2, thick]
				coordinates {
					(0, 0.000000)
					(1, 0.000000)
					(2, 0.000000)
					(3, 0.001953)
					(4, 0.001953)
					(5, 0.001953)
					(6, 0.001953)
					(7, 0.003906)
					(8, 0.003906)
					(9, 0.005859)
					(10, 0.007812)
					(11, 0.007812)
					(12, 0.011719)
					(13, 0.011719)
					(14, 0.013672)
					(15, 0.013672)
					(16, 0.015625)
					(17, 0.017578)
					(18, 0.019531)
					(19, 0.019531)
					(20, 0.021484)
					(21, 0.023438)
					(22, 0.025391)
					(23, 0.025391)
					(24, 0.027344)
					(25, 0.025391)
					(26, 0.031250)
					(27, 0.029297)
					(28, 0.029297)
					(29, 0.031250)
					(30, 0.029297)
					(31, 0.031250)
					(32, 0.029297)
					(33, 0.033203)
					(34, 0.031250)
					(35, 0.029297)
					(36, 0.031250)
					(37, 0.025391)
					(38, 0.027344)
					(39, 0.027344)
					(40, 0.025391)
					(41, 0.027344)
					(42, 0.025391)
					(43, 0.021484)
					(44, 0.019531)
					(45, 0.021484)
					(46, 0.017578)
					(47, 0.017578)
					(48, 0.015625)
					(49, 0.013672)
					(50, 0.011719)
					(51, 0.011719)
					(52, 0.009766)
					(53, 0.007812)
					(54, 0.007812)
					(55, 0.005859)
					(56, 0.003906)
					(57, 0.003906)
					(58, 0.001953)
					(59, 0.001953)
					(60, 0.001953)
					(61, 0.001953)
					(62, 0.000000)
					(63, 0.000000)
				};
				\addlegendentry{\textcolor{HighlightColor2}{sample density}}
			\end{axis}
		\end{tikzpicture}
		\caption{Alias Method, $2^{9}$ samples}
	\end{subfigure}
	\begin{subfigure}{0.48\textwidth}
		\centering
		\begin{tikzpicture}
			\begin{axis}[
				ticks=none,
				legend style={draw=none},
				legend style={at={(0.5,0.05)},anchor=south, fill=none, font=\tiny},
				legend cell align=left,
				width=\textwidth,
				height=.3\textheight]
				\addlegendimage{empty legend}
				\addlegendentry{}
				\addplot[color=black, thick]
				coordinates {
					(0, 0.000000)
					(1, 0.000075)
					(2, 0.000300)
					(3, 0.000673)
					(4, 0.001189)
					(5, 0.001845)
					(6, 0.002633)
					(7, 0.003547)
					(8, 0.004576)
					(9, 0.005713)
					(10, 0.006944)
					(11, 0.008259)
					(12, 0.009646)
					(13, 0.011089)
					(14, 0.012577)
					(15, 0.014093)
					(16, 0.015625)
					(17, 0.017157)
					(18, 0.018673)
					(19, 0.020161)
					(20, 0.021604)
					(21, 0.022991)
					(22, 0.024306)
					(23, 0.025537)
					(24, 0.026674)
					(25, 0.027703)
					(26, 0.028617)
					(27, 0.029405)
					(28, 0.030061)
					(29, 0.030577)
					(30, 0.030950)
					(31, 0.031175)
					(32, 0.031250)
					(33, 0.031175)
					(34, 0.030950)
					(35, 0.030577)
					(36, 0.030061)
					(37, 0.029405)
					(38, 0.028617)
					(39, 0.027703)
					(40, 0.026674)
					(41, 0.025537)
					(42, 0.024306)
					(43, 0.022991)
					(44, 0.021604)
					(45, 0.020161)
					(46, 0.018673)
					(47, 0.017157)
					(48, 0.015625)
					(49, 0.014093)
					(50, 0.012577)
					(51, 0.011089)
					(52, 0.009646)
					(53, 0.008259)
					(54, 0.006944)
					(55, 0.005713)
					(56, 0.004576)
					(57, 0.003547)
					(58, 0.002633)
					(59, 0.001845)
					(60, 0.001189)
					(61, 0.000673)
					(62, 0.000300)
					(63, 0.000075)
				};
				\addlegendentry{\textcolor{black}{target density}}

				\addplot[color=HighlightColor, thick]
				coordinates {
					(0, 0.000000)
					(1, 0.001953)
					(2, 0.000000)
					(3, 0.000000)
					(4, 0.001953)
					(5, 0.001953)
					(6, 0.001953)
					(7, 0.003906)
					(8, 0.003906)
					(9, 0.005859)
					(10, 0.007812)
					(11, 0.007812)
					(12, 0.009766)
					(13, 0.009766)
					(14, 0.013672)
					(15, 0.013672)
					(16, 0.015625)
					(17, 0.017578)
					(18, 0.017578)
					(19, 0.021484)
					(20, 0.021484)
					(21, 0.023438)
					(22, 0.023438)
					(23, 0.025391)
					(24, 0.027344)
					(25, 0.027344)
					(26, 0.029297)
					(27, 0.029297)
					(28, 0.029297)
					(29, 0.031250)
					(30, 0.031250)
					(31, 0.031250)
					(32, 0.029297)
					(33, 0.031250)
					(34, 0.031250)
					(35, 0.031250)
					(36, 0.029297)
					(37, 0.029297)
					(38, 0.029297)
					(39, 0.027344)
					(40, 0.027344)
					(41, 0.025391)
					(42, 0.023438)
					(43, 0.023438)
					(44, 0.021484)
					(45, 0.021484)
					(46, 0.017578)
					(47, 0.017578)
					(48, 0.015625)
					(49, 0.013672)
					(50, 0.013672)
					(51, 0.009766)
					(52, 0.009766)
					(53, 0.007812)
					(54, 0.007812)
					(55, 0.005859)
					(56, 0.003906)
					(57, 0.003906)
					(58, 0.001953)
					(59, 0.001953)
					(60, 0.001953)
					(61, 0.000000)
					(62, 0.000000)
					(63, 0.000000)
				};
				\addlegendentry{\textcolor{HighlightColor}{sample density}}
			\end{axis}
		\end{tikzpicture}
		\caption{monotonic mapping, $2^{9}$ samples}
		\end{subfigure}\hfill\\\vspace*{1em}\\
	\begin{subfigure}{0.48\textwidth}
		\centering
		\begin{tikzpicture}
			\begin{axis}[
				ticks=none,
				legend style={draw=none},
				legend style={at={(0.5,0.05)},anchor=south, fill=none, font=\tiny},
				legend cell align=left,
				width=\textwidth,
				height=.3\textheight]
				\addlegendimage{empty legend}
				\addlegendentry{}
				\addplot[color=black, name path=f, thick]
				coordinates {
					(0, 0.000000)
					(1, 0.000075)
					(2, 0.000300)
					(3, 0.000673)
					(4, 0.001189)
					(5, 0.001845)
					(6, 0.002633)
					(7, 0.003547)
					(8, 0.004576)
					(9, 0.005713)
					(10, 0.006944)
					(11, 0.008259)
					(12, 0.009646)
					(13, 0.011089)
					(14, 0.012577)
					(15, 0.014093)
					(16, 0.015625)
					(17, 0.017157)
					(18, 0.018673)
					(19, 0.020161)
					(20, 0.021604)
					(21, 0.022991)
					(22, 0.024306)
					(23, 0.025537)
					(24, 0.026674)
					(25, 0.027703)
					(26, 0.028617)
					(27, 0.029405)
					(28, 0.030061)
					(29, 0.030577)
					(30, 0.030950)
					(31, 0.031175)
					(32, 0.031250)
					(33, 0.031175)
					(34, 0.030950)
					(35, 0.030577)
					(36, 0.030061)
					(37, 0.029405)
					(38, 0.028617)
					(39, 0.027703)
					(40, 0.026674)
					(41, 0.025537)
					(42, 0.024306)
					(43, 0.022991)
					(44, 0.021604)
					(45, 0.020161)
					(46, 0.018673)
					(47, 0.017157)
					(48, 0.015625)
					(49, 0.014093)
					(50, 0.012577)
					(51, 0.011089)
					(52, 0.009646)
					(53, 0.008259)
					(54, 0.006944)
					(55, 0.005713)
					(56, 0.004576)
					(57, 0.003547)
					(58, 0.002633)
					(59, 0.001845)
					(60, 0.001189)
					(61, 0.000673)
					(62, 0.000300)
					(63, 0.000075)
				};
				\addlegendentry{\textcolor{black}{target density}}

				\addplot[color=HighlightColor2, thick]
				coordinates {
					(0, 0.000000)
					(1, 0.000000)
					(2, 0.000488)
					(3, 0.000488)
					(4, 0.000977)
					(5, 0.001953)
					(6, 0.002441)
					(7, 0.004395)
					(8, 0.004395)
					(9, 0.005859)
					(10, 0.006836)
					(11, 0.008301)
					(12, 0.009766)
					(13, 0.010742)
					(14, 0.012207)
					(15, 0.014648)
					(16, 0.015625)
					(17, 0.017578)
					(18, 0.018555)
					(19, 0.020020)
					(20, 0.021484)
					(21, 0.022949)
					(22, 0.025391)
					(23, 0.025391)
					(24, 0.026855)
					(25, 0.027832)
					(26, 0.027832)
					(27, 0.028809)
					(28, 0.030762)
					(29, 0.030273)
					(30, 0.030762)
					(31, 0.031250)
					(32, 0.031250)
					(33, 0.031738)
					(34, 0.030762)
					(35, 0.030273)
					(36, 0.030273)
					(37, 0.028809)
					(38, 0.029297)
					(39, 0.028320)
					(40, 0.025879)
					(41, 0.025879)
					(42, 0.024414)
					(43, 0.023438)
					(44, 0.020996)
					(45, 0.020020)
					(46, 0.019043)
					(47, 0.016602)
					(48, 0.015625)
					(49, 0.014160)
					(50, 0.012695)
					(51, 0.011230)
					(52, 0.009277)
					(53, 0.007812)
					(54, 0.007324)
					(55, 0.005859)
					(56, 0.004395)
					(57, 0.003418)
					(58, 0.002441)
					(59, 0.001953)
					(60, 0.000977)
					(61, 0.000977)
					(62, 0.000000)
					(63, 0.000000)
				};
				\addlegendentry{\textcolor{HighlightColor2}{sample density}}
			\end{axis}
		\end{tikzpicture}
		\caption{Alias Method, $2^{11}$ samples}
	\end{subfigure}
	\begin{subfigure}{0.48\textwidth}
		\centering
		\begin{tikzpicture}
			\begin{axis}[
				ticks=none,
				legend style={draw=none},
				legend style={at={(0.5,0.05)},anchor=south, fill=none, font=\tiny},
				legend cell align=left,
				width=\textwidth,
				height=.3\textheight]
				\addlegendimage{empty legend}
				\addlegendentry{}
				\addplot[color=black, thick]
				coordinates {
					(0, 0.000000)
					(1, 0.000075)
					(2, 0.000300)
					(3, 0.000673)
					(4, 0.001189)
					(5, 0.001845)
					(6, 0.002633)
					(7, 0.003547)
					(8, 0.004576)
					(9, 0.005713)
					(10, 0.006944)
					(11, 0.008259)
					(12, 0.009646)
					(13, 0.011089)
					(14, 0.012577)
					(15, 0.014093)
					(16, 0.015625)
					(17, 0.017157)
					(18, 0.018673)
					(19, 0.020161)
					(20, 0.021604)
					(21, 0.022991)
					(22, 0.024306)
					(23, 0.025537)
					(24, 0.026674)
					(25, 0.027703)
					(26, 0.028617)
					(27, 0.029405)
					(28, 0.030061)
					(29, 0.030577)
					(30, 0.030950)
					(31, 0.031175)
					(32, 0.031250)
					(33, 0.031175)
					(34, 0.030950)
					(35, 0.030577)
					(36, 0.030061)
					(37, 0.029405)
					(38, 0.028617)
					(39, 0.027703)
					(40, 0.026674)
					(41, 0.025537)
					(42, 0.024306)
					(43, 0.022991)
					(44, 0.021604)
					(45, 0.020161)
					(46, 0.018673)
					(47, 0.017157)
					(48, 0.015625)
					(49, 0.014093)
					(50, 0.012577)
					(51, 0.011089)
					(52, 0.009646)
					(53, 0.008259)
					(54, 0.006944)
					(55, 0.005713)
					(56, 0.004576)
					(57, 0.003547)
					(58, 0.002633)
					(59, 0.001845)
					(60, 0.001189)
					(61, 0.000673)
					(62, 0.000300)
					(63, 0.000075)
				};
				\addlegendentry{\textcolor{black}{target density}}

				\addplot[color=HighlightColor, thick]
				coordinates {
					(0, 0.000000)
					(1, 0.000488)
					(2, 0.000000)
					(3, 0.000977)
					(4, 0.000977)
					(5, 0.001953)
					(6, 0.002441)
					(7, 0.003906)
					(8, 0.004395)
					(9, 0.005859)
					(10, 0.006836)
					(11, 0.008301)
					(12, 0.009277)
					(13, 0.011230)
					(14, 0.012695)
					(15, 0.014160)
					(16, 0.015625)
					(17, 0.017090)
					(18, 0.018555)
					(19, 0.020020)
					(20, 0.021973)
					(21, 0.022949)
					(22, 0.024414)
					(23, 0.025391)
					(24, 0.026855)
					(25, 0.027344)
					(26, 0.028809)
					(27, 0.029297)
					(28, 0.030273)
					(29, 0.030273)
					(30, 0.031250)
					(31, 0.031250)
					(32, 0.030762)
					(33, 0.031250)
					(34, 0.031250)
					(35, 0.030273)
					(36, 0.030273)
					(37, 0.029297)
					(38, 0.028809)
					(39, 0.027344)
					(40, 0.026855)
					(41, 0.025391)
					(42, 0.024414)
					(43, 0.022949)
					(44, 0.021973)
					(45, 0.020020)
					(46, 0.018555)
					(47, 0.017090)
					(48, 0.015625)
					(49, 0.014160)
					(50, 0.012695)
					(51, 0.011230)
					(52, 0.009277)
					(53, 0.008301)
					(54, 0.006836)
					(55, 0.005859)
					(56, 0.004395)
					(57, 0.003906)
					(58, 0.002441)
					(59, 0.001953)
					(60, 0.000977)
					(61, 0.000977)
					(62, 0.000000)
					(63, 0.000000)
				};
				\addlegendentry{\textcolor{HighlightColor}{sample density}}
			\end{axis}
		\end{tikzpicture}
		\caption{monotonic mapping, $2^{11}$ samples}
	\end{subfigure}
		\caption{Sampling proportional to a one-dimensional density with the Alias Method often converges significantly slower
than sampling the inverse cumulative distribution function, especially in regions with high densities. In this example the discrete target
probability distribution function is the continuous black curve sampled at 64 equidistant steps.}
		\label{Fig:results_sine64}
\end{figure}

Reordering the intervals creates a different, non-monotonic mapping (see \Cref{Fig:alias_map_details}), which comes with unpleasant side effects
for quasi-Monte Carlo methods \cite{Nie:92}. As intervals are reordered, the low discrepancy of a sequence may be destroyed (see \Cref{Fig:Teaser}).
This especially affects regions with high probabilities, which is apparent in the one dimensional example in \Cref{Fig:results_sine64}.

In such regions, many samples are aliases of samples in low-density regions, which is intrinsic to the
construction of the Alias Method. Hence the resulting set of samples cannot be guaranteed to
be of low discrepancy, which may harm convergence speed.%

\begin{figure}
	\begin{tabular}[t]{@{}c@{}}
	\begin{tikzpicture}
		\node[inner sep=0,outer sep=0] at (0, 0) {\includegraphics[width=\textwidth, trim={0, 20, 0, 20}, clip]{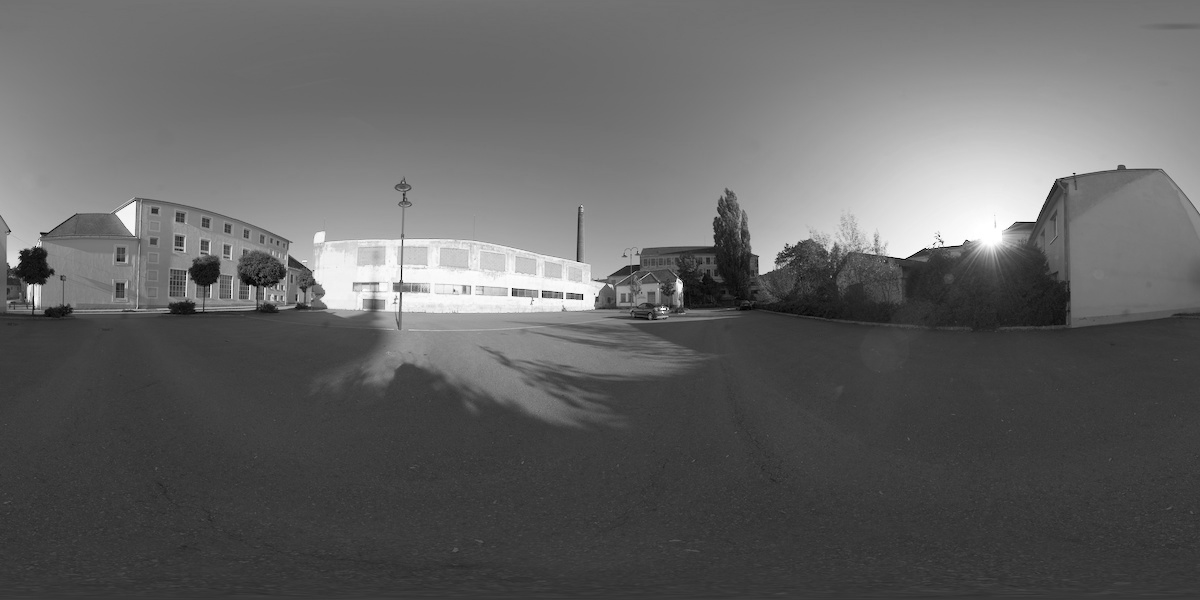}};
		\draw[GoogleColor] (2.95, 0.4) rectangle (4.2, 1.2);
	\end{tikzpicture}\\
	(a) target density
	\end{tabular}
	\setlength\tabcolsep{0 pt}
	\begin{tabular}[t]{@{}cc@{}}
		\includegraphics[width=.5\textwidth, trim={200 200 200 200}, clip]{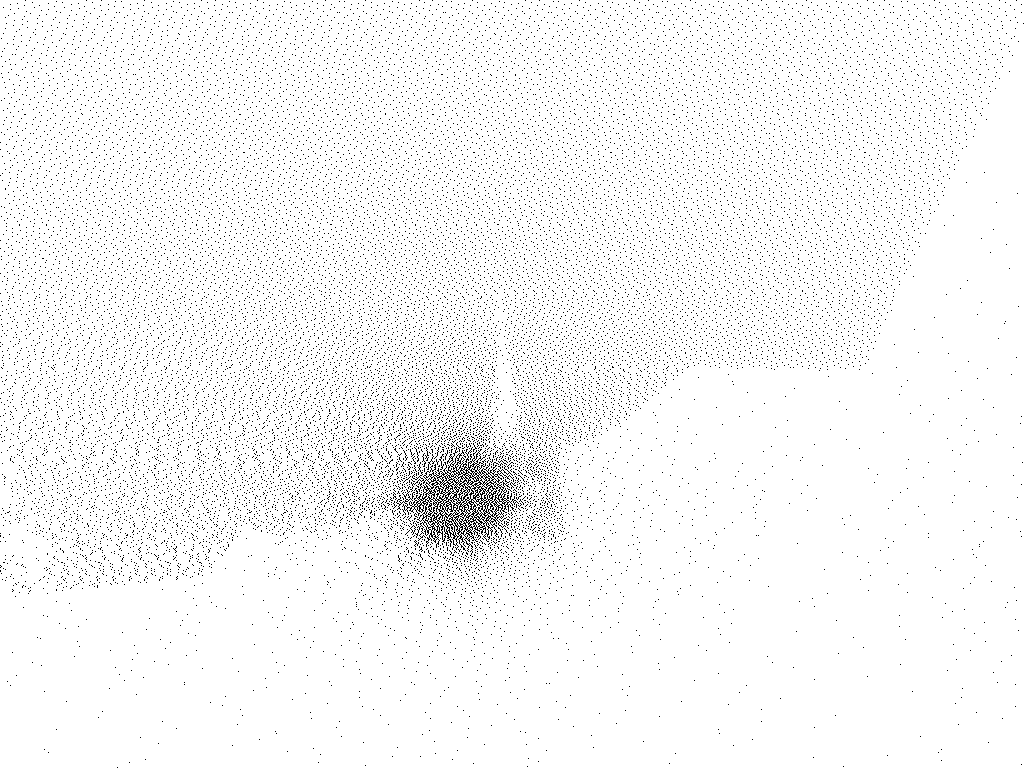}&
		\includegraphics[width=.5\textwidth, trim={200 200 200 200}, clip]{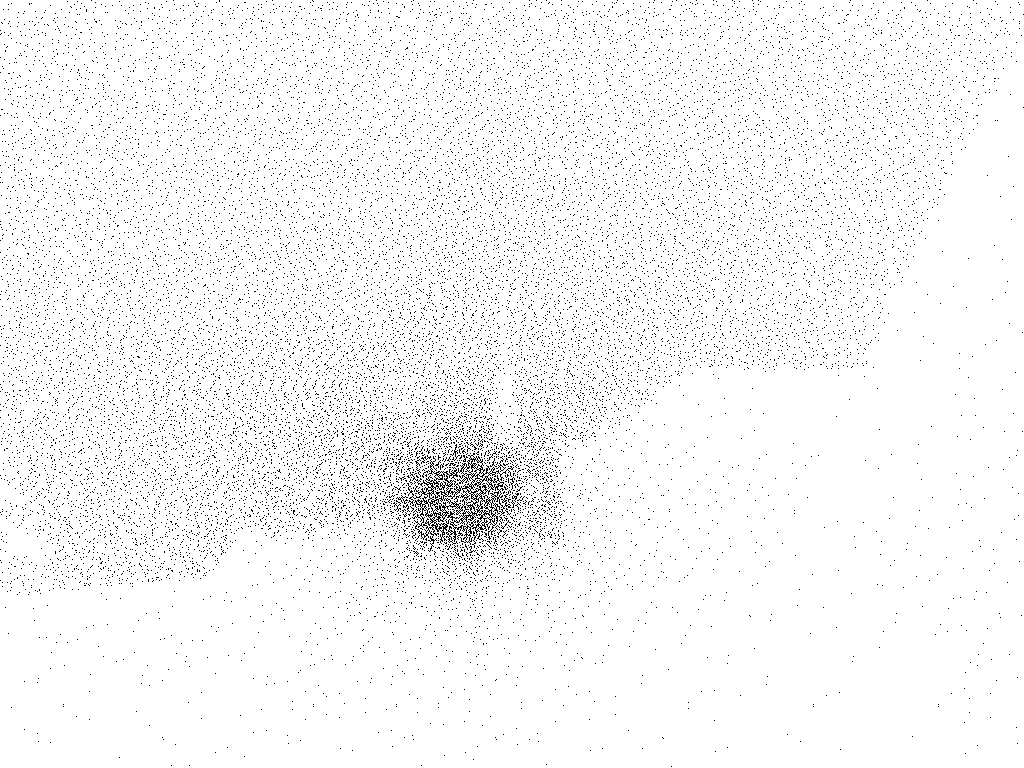}\\
		(b) monotonic\ mapping & (c) Alias Method
	\end{tabular}
	\caption{Sampling the two-dimensional target density in (a) using the two-dimensional low discrepancy Hammersley point
	set as a sequence of uniformly distributed points by (b) transforming them with a monotonic mapping preserves the
	distribution properties in a warped space, whereas (c) transforming with the Alias Method degrades the uniformity of
	the distribution. This is especially bad in regions with a high density. Image copyright openfootage.net.}%
	\label{Fig:2d_zoom}
\end{figure}

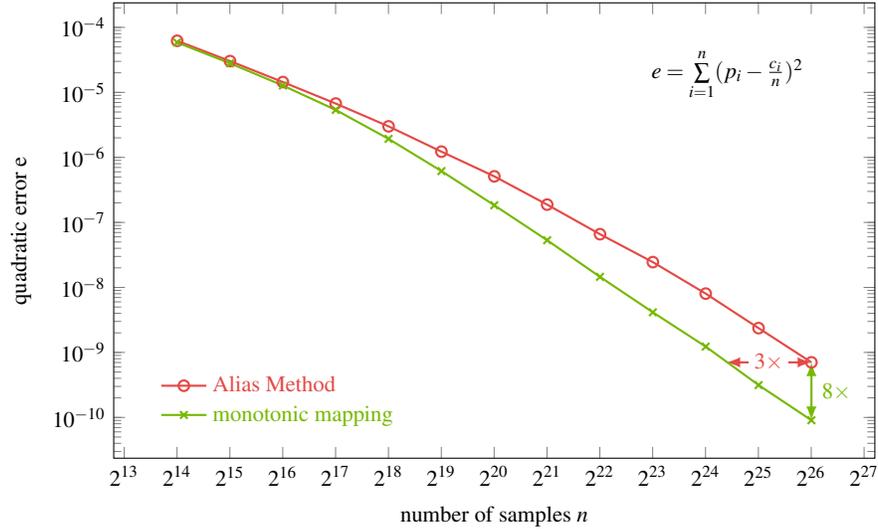
\begin{figure}[h]
	\begin{tikzpicture}
		\begin{axis}[
			ylabel={quadratic error e},
			xlabel=number of samples $n$,
			legend style={draw=none},
			legend style={at={(0.05,0.05)},anchor=south west, fill=none, font=\small},
			ymode=log,
			xmode=log,
			log basis x={2},
			legend cell align=left,
			width=\textwidth,
			height=.4\textheight]
			\addlegendimage{empty legend}
			\addlegendentry{}
			\draw[GoogleColor, latex-latex, thick] (axis cs: 67108864, 7.03782e-10) -- (axis cs: 22000000, 7.03782e-10) node [pos=0.5, fill=white, font=\small, inner sep=2, outer sep=0] {$3 \times$};
			\draw[HighlightColor, latex-latex, thick] (axis cs: 67108864, 7.03782e-10) -- (axis cs: 67108864, 9.05748e-11) node [right,pos=0.5, font=\small, inner sep=2, outer sep=2] {$8 \times$};
			\node[anchor=north east] at (axis cs: 67108864, 6.21921e-05) {$e = \sum_{i = 1}^{n}\limits(p_i - \frac{c_i}{n})^2$};
			\addplot[color=GoogleColor,mark=o, thick]
			coordinates {
				(16384, 6.21921e-05)
				(32768, 3.03133e-05)
				(65536, 1.44225e-05)
				(131072, 6.73274e-06)
				(262144, 3.00216e-06)
				(524288, 1.22591e-06)
				(1048576, 5.11157e-07)
				(2097152, 1.87826e-07)
				(4194304, 6.58932e-08)
				(8388608, 2.45068e-08)
				(16777216, 8.02817e-09)
				(33554432, 2.36566e-09)
				(67108864, 7.03782e-10)
			};
			\addlegendentry{\textcolor{GoogleColor}{Alias Method}}
			\addplot[color=HighlightColor,mark=x, thick]
			coordinates {
				(16384, 5.85265e-05)
				(32768, 2.8031e-05)
				(65536, 1.27524e-05)
				(131072, 5.3703e-06)
				(262144, 1.92788e-06)
				(524288, 6.16251e-07)
				(1048576, 1.8354e-07)
				(2097152, 5.32163e-08)
				(4194304, 1.45417e-08)
				(8388608, 4.14151e-09)
				(16777216, 1.22913e-09)
				(33554432, 3.1493e-10)
				(67108864, 9.05748e-11)
			};
			\addlegendentry{\textcolor{HighlightColor}{monotonic mapping}}
		\end{axis}
	\end{tikzpicture}
	\caption{As the Alias Method does not preserve the low discrepancy of the sequence, convergence speed suffers. 
	Sampling the two-dimensional density detailed in \Cref{Fig:2d_zoom} with the Alias Method, the quadratic error for $2^{26}$ samples is $8\times$ as high, and $3x$ as many samples are required to get the
	same error as sampling with the inverse mapping. $n$ is the total number of samples, while $c_i$ is the number of
	samples that realized the value $i$.} %
	\label{Fig:convergence}
\end{figure}

A typical application in graphics is sampling according to a two-dimensional density map (``target density''). \Cref{Fig:2d_zoom} shows two
regions of such a map with points sampled using the Alias Method and the inverse mapping. Points are generated by first
selecting a row according to the density distribution of the rows and then selecting the column according to the
distribution in this row. Already a visual comparison identifies the shortcomings of the Alias Method, which the
evaluation of the quadratic deviation of the sampled density to the target density in \Cref{Fig:convergence} confirms.

Known algorithms for setting up the necessary data structures of the Alias Method are serial with a run time considerably higher than
the prefix sum required for inversion methods, which can be efficiently calculated in parallel. %

\section{An Efficient Inverse Mapping} \label{Sec:Method} %

In the following, a method is introduced that accelerates finding the inverse $P^{-1}$ of the cumulative distribution function and preserves
distribution properties, such as the star-discrepancy. Note that we aim to  preserve these properties in the warped space in which
regions with a high density are stretched, and regions with a low density are squashed. Unwarping the space like in
\Cref{Fig:Teaser} reveals that the sequence in warped space is just partitioned into these regions.
Similar to the Cutpoint Method with binary search, the mapping can be performed in $\mathcal{O}(1)$ on the average and
in $\mathcal{O}(\log_2 n)$ in the worst case. However, we improve the average case performance by explicitly storing a
more optimal hierarchical structure that improves the average case of finding values that cannot be immediately
identified using the guide table. While for degenerate hierarchical structures the worst case may increase to
$\mathcal{O}(n)$, a fallback method constructs such a structure upon detection to guarantee logarithmic complexity.

Instead of optimizing for the minimal required memory footprint to guarantee access in constant time, we dedicate
$\mathcal{O}(n)$ additional space for the hierarchical structure and propose to either assign as much memory as
affordable for the guide table or to select its size proportional to the size of the discrete probability distribution
$p$.

As explained in the previous section, using the monotonic mapping $P^{-1}$, and thus avoiding the Alias Method, is crucial for 
efficient quasi-Monte Carlo integration. At the same time, the massively parallel execution on Graphics Processing Units (GPUs)
suffers more from outliers in the execution time than serial computation since computation is performed by a number of threads organized in
groups (``warps''), which need to synchronize and hence only finish after the last thread in this group has terminated.
Therefore, lengthy computations that would otherwise be averaged out need to be avoided.

Note that the Cutpoint Method uses the monotonic mapping $P^{-1}$ and with binary search avoids the worst case. However, it does not yield
good performance for the majority of cases in which an additional search in each cell needs to be performed. In what follows,
we therefore combine the Cutpoint Method with a binary tree to especially optimize these cases.

In that context, radix trees are of special interest since they can be very efficiently built in parallel \cite{ParallelLBVH,ParallelKD}. \Cref{Sec:Radix_trees}
introduces their properties and their efficient parallel construction. Furthermore, their underlying structure that
splits intervals in the middle is nearly optimal for this application.

However, as many trees of completely different sizes are required, a na\"ive implementation that builds these trees in
parallel results in severe load balancing issues. In \Cref{Sec:Radix_forests} we therefore introduce a method that
builds the entire radix tree forest simultaneously in parallel, but instead of parallelizing over trees, parallelization is
uniformly distributed over the data.

\subsection{Massively Parallel Construction of Radix Trees}
\label{Sec:Radix_trees}

In a radix tree, also often called compact prefix tree, the value of each leaf node is the concatenation of the values
on the path to it from the root of the tree. The number of children of each internal node is always greater than one;
otherwise the values of the node and its child can be concatenated already in the node itself.

The input values referenced in the leaves of such a tree must be strictly ordered, which for arbitrary data requires an
additional sorting step and an indirection from the index $i'$ of the value in the sorted input data to the original
index $i$. For the application of such a tree to perform the inverse mapping, i.e. to identify the interval $m$ that includes $\xi$, the
input values are the lower bounds of the intervals, which by construction are already sorted.

\begin{figure}
	\includegraphics[width=\linewidth]{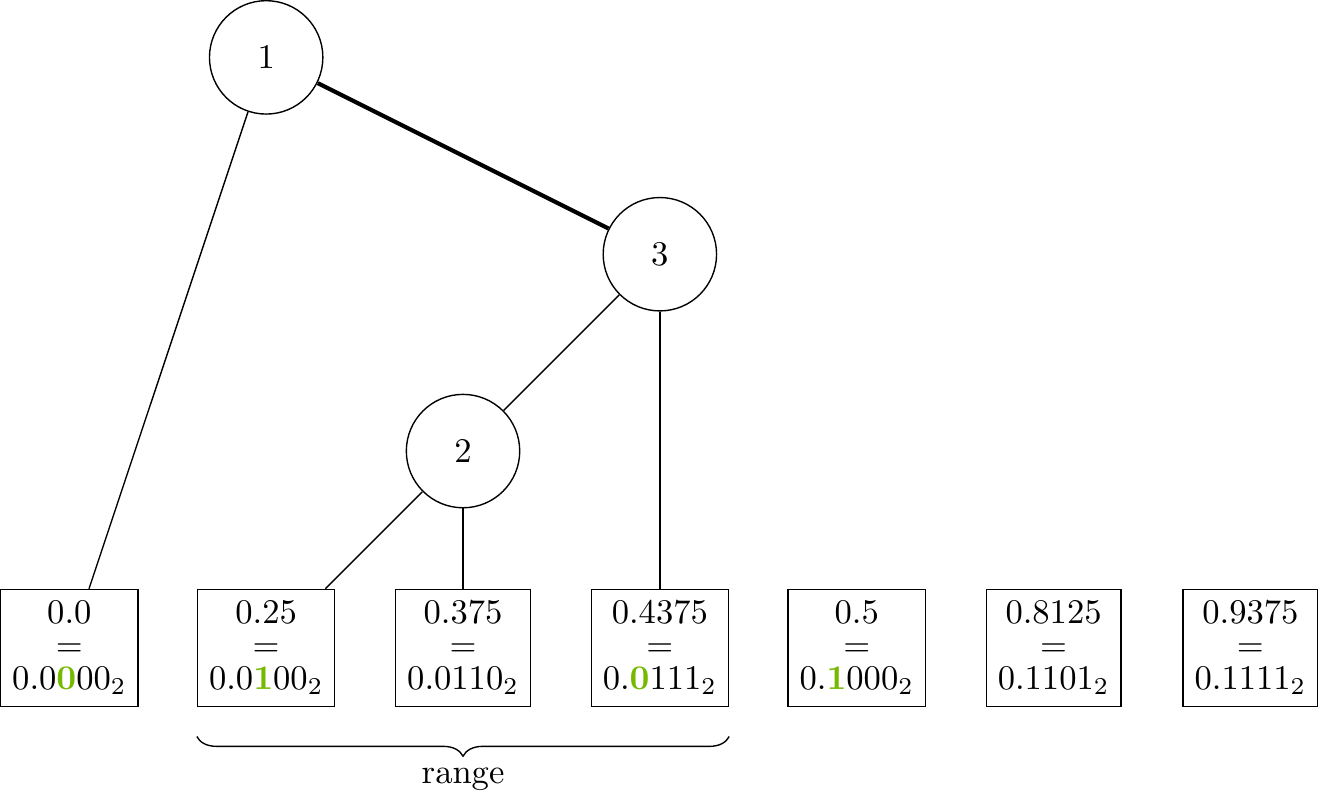}
	\caption{In each merging step of the bottom-up construction method for the radix tree, the value of the leftmost leaf in
	the range of leaf nodes below the current root is compared to the value of its left neighbor, and the value of the rightmost leaf in the
	range is compared to the value of its right neighbor. In both comparisons, the most significant differing bit of
	the value in base two is determined. Since the most significant differing bit of the comparison on the left side of the
	current range is less significant than the most significant differing bit on the right side, the root node of the
	current subtree (here 3) is a right child, and the index of its parent node equals to the lowest index of the leaves
	in the range (here 1).
	}
	\label{Fig:radix_tree_merge}
\end{figure}

Radix trees over integers with a radix of two are of particular interest for hardware and software operating on binary
numbers. Since by definition each internal node has exactly two children, there exists an enumeration scheme for
these trees that determines the index of each node only given the range of input values below its children \cite{ParallelLBVH}: The index of
each node is the lowest data index below its right, or,
equivalently, the highest data index below its left child plus one. This information is not only available if the tree is built
top-down by recursive bisection, but can also easily be propagated up when building the tree bottom-up.
The index of the parent node, implied by this enumeration rule, is then the index of the leftmost node in the range of
leaf nodes below the current root node if the node is a right child, and equals the index of the rightmost node in the
range plus on if it is a left child.

Determining whether a node is a left
or a right child can be done by comparing the values in the leaves at the boundaries of the range of nodes below the
current root node to their neighbors outside the range. If the binary distance, determined by taking the bit-wise \emph{exclusive or},
of the value of the leftmost leaf in the range to those of its left neighbor is smaller than the
distance of the value of the rightmost leaf node in the range to those of its right neighbor, it must be a right child. Otherwise it is
a left child. In case the leftmost leaf in the range is the first one, or the rightmost leaf in the range is the last
one, a comparison to the neighbor is not possible, and these nodes must be a left and a right child, respectively. An example of such a decision is shown in \Cref{Fig:radix_tree_merge}.

Given this scheme, these trees can be built from bottom up completely in parallel since all information required to
perform one merging step can be retrieved without synchronization. Like in the parallel construction of Radix Trees by
Apetrei \cite{ParallelLBVH}, the only synchronization required ensures that only
one of the children merges further up, and does that after its sibling has already reported the range of nodes below it.

The implementation of the method presented in
\Cref{alg:radix_forest} uses an atomic exchange operation
(\emph{atomicExch}) and an array initialized to -1 for synchronization. The $\oplus$ operator performs a bitwise
\emph{exclusive or} operation on the floating point representation of the input value. As the ordering of IEEE 754 floating point
numbers is equal to the binary ordering, taking the \emph{exclusive or} of two floating point numbers
calculates a value in which the position of the most significant bit set to one indicates the highest level in the implicit tree
induced by recursive bisection of the interval $[0, 1)$ on which the two values are not referenced by the same child.
Hence, the bitwise \emph{exclusive or} operation determines the distance of two values in such a tree.

\subsection{Massively Parallel Construction of Radix Tree Forests}
\label{Sec:Radix_forests}

A forest of radix trees can be built in a similar way. Again, parallelization runs over the whole data range, not over
the individual trees in the forest. Therefore, perfect load balancing can be achieved. As highlighted in \Cref{alg:radix_forest},
the key difference between building a single tree and a forest
is that in each merging step it must be checked whether merging would
go over partition boundaries of the forest. Avoiding such a merge operation over a boundary is as simple as setting the
distance (again computed using $\oplus$) to the maximum.

Note that node indices for small (sub-) trees are always consecutive by design, which improves cache hit rates, which
furthermore can be improved by interleaving the values of the cumulative distribution function and the indices of the children.

\begin{algorithm}
	\begin{algorithmic}
		\STATE \textbf{Input:} $data \in \left[0, 1\right)^n$ in increasing order, {\color{HighlightColor} number of partitions $m$}
		\STATE \textbf{Output:} $n\ nodes$, each with indices of the left and right child
		\STATE $otherBounds \gets (-1, ..., -1) \in \mathbb{Z}^n$
		\STATE $data[-1] \gets data[n] = 1$
		\FOR{$i \in \left[ 0, n \right)$ \textbf{in parallel}}
			\STATE $nodeId \gets leaf_i$
			\STATE {\color{HighlightColor} $curCell \gets \lfloor data[i] \cdot m \rfloor$}
			\STATE $range \gets \left(i, i\right)$
			\REPEAT
			\STATE $valueLow \gets data[range[0]]$
			\STATE $valueHigh \gets data[range[1]]$
			\STATE $valueNeighborLow \gets data[range[0] - 1]$
			\STATE $valueNeighborHigh \gets data[range[1] + 1]$
			{\color{HighlightColor} \IF{$\lfloor valueNeighborLow \cdot m \rfloor < curCell$}
				\STATE $valueNeighborLow \gets 1$
			\ENDIF
			\IF{$\lfloor valueNeighborHigh \cdot m \rfloor > curCell$}
				\STATE $valueNeighborHigh \gets 1$
			\ENDIF}
			\STATE $child \gets \begin{cases}0 & valueLow \oplus valueNeighborLow > valueHigh \oplus valueNeighborHigh\\ 1 & \textrm{otherwise}\end{cases}$
			\STATE $parent \gets \begin{cases}range[1] + 1 & child = 0\\range[0] & child = 1\end{cases}$
			\STATE $nodes[parent].child[child] \gets nodeId$
			\STATE $otherBound \gets \textbf{atomicExch}\left(otherBounds[parent], range[child]\right)$ %
			\IF{$otherBound \neq -1$}
				\STATE $range[1 - child] \gets otherBound$
				\STATE $nodeId \gets parent$
			\ENDIF
			\UNTIL{otherBound = -1}
		\ENDFOR
	\end{algorithmic}
	\caption{Parallel constructing a radix tree forest. Omitting the colored parts results in the construction of a radix
	tree, only (see \cite{ParallelLBVH}).}
	\label{alg:radix_forest}
\end{algorithm}

\begin{figure}
	\begin{minicg}
	import math
	probs_construction = [math.ceil((0.12 * random.random() if x < 7 else random.random()) * 64) / 64.0 for x in range(12)]
	cdf_construction = []
	cur = 0
	for p in probs_construction:
			cur += p
			cdf_construction.append(cur)

	for i in range(len(probs_construction)):
			probs_construction[i] /= cur
			cdf_construction[i] /= cur

	Tikz.begin()
	xscale = 10.7
	yscale = 0.8
	xoffset = -0.5
	yoffset = -1.5
	tree = RadixTree()
	tree.build(probs_construction, False, xscale, yscale, False, False, '', 0.0, True, colors)

	old_bin = 0
	start = 0
	xscale += 1 # because of -0.5 and +0.5
	for i, c in enumerate(cdf_construction):
			box_start = (xscale * start + xoffset, -yscale + yoffset + 1 - 0.5)
			Tikz.fill(box_start, colors[mod(i, len(colors))]).rectangle((xscale * c + xoffset, -yscale + yoffset - 0.5))
			Tikz.node_with_label(str(i), "b{}".format(i), {'white':'', 'pos': 0.5, 'font': '\\tiny'}).endl()
			start = c

	Tikz.draw((xoffset, -yscale + yoffset + 1.1 - 0.5), 'thick, fill=black, fill opacity=0.1').rectangle((xoffset + xscale, -yscale + yoffset - 0.1 - 0.5)).endl()
	for i in range(len(cdf_construction)):
			Tikz.draw((float(i) / len(cdf_construction) * xscale + xoffset, -yscale + yoffset + 1.1 - 0.5), 'thick, black').to((float(i) / len(cdf_construction) * xscale + xoffset, -yscale + yoffset - 0.1 - 0.5)).endl()

	Tikz.end()
	\end{minicg}
	\begin{tikzpicture}[]
% show construction = False
% show construction = False
\fill[fill opacity=0.1, black] (-0.48636363636363633, -0.95) rectangle (0.486363636364, 2.85) ;
\node[font=\tiny, anchor=south] at (0.0, -0.7) { 0 } ;
\fill[fill opacity=0.05, black] (0.48636363636363633, -0.95) rectangle (1.45909090909, 2.85) ;
\node[font=\tiny, anchor=south] at (0.972727272727, -0.7) { 1 } ;
\fill[fill opacity=0.1, black] (1.459090909090909, -0.95) rectangle (2.43181818182, 2.85) ;
\node[font=\tiny, anchor=south] at (1.94545454545, -0.7) { 2 } ;
\fill[fill opacity=0.05, black] (2.4318181818181817, -0.95) rectangle (3.40454545455, 2.85) ;
\node[font=\tiny, anchor=south] at (2.91818181818, -0.7) { 3 } ;
\fill[fill opacity=0.1, black] (3.4045454545454543, -0.95) rectangle (4.37727272727, 2.85) ;
\node[font=\tiny, anchor=south] at (3.89090909091, -0.7) { 4 } ;
\fill[fill opacity=0.05, black] (4.377272727272727, -0.95) rectangle (5.35, 2.85) ;
\node[font=\tiny, anchor=south] at (4.86363636364, -0.7) { 5 } ;
\fill[fill opacity=0.1, black] (5.35, -0.95) rectangle (6.32272727273, 2.85) ;
\node[font=\tiny, anchor=south] at (5.83636363636, -0.7) { 6 } ;
\fill[fill opacity=0.05, black] (6.322727272727272, -0.95) rectangle (7.29545454545, 2.85) ;
\node[font=\tiny, anchor=south] at (6.80909090909, -0.7) { 7 } ;
\fill[fill opacity=0.1, black] (7.295454545454544, -0.95) rectangle (8.26818181818, 2.85) ;
\node[font=\tiny, anchor=south] at (7.78181818182, -0.7) { 8 } ;
\fill[fill opacity=0.05, black] (8.268181818181818, -0.95) rectangle (9.24090909091, 2.85) ;
\node[font=\tiny, anchor=south] at (8.75454545455, -0.7) { 9 } ;
\fill[fill opacity=0.1, black] (9.24090909090909, -0.95) rectangle (10.2136363636, 2.85) ;
\node[font=\tiny, anchor=south] at (9.72727272727, -0.7) { 10 } ;
\fill[fill opacity=0.05, black] (10.213636363636363, -0.95) rectangle (11.1863636364, 2.85) ;
\node[font=\tiny, anchor=south] at (10.7, -0.7) { 11 } ;
\node[draw=black, font=\tiny, inner sep=2, fill=HighlightColor] (l0) at (0.0, 0) { 0.00000 } ;
\node[circle, font=\tiny, draw=black, inner sep=2] (1) at (0.972727272727, 1.6) { 1 } ;
\draw[] (1) -- (l0) ;
\node[draw=black, font=\tiny, inner sep=2, fill=HighlightColor2] (l1) at (0.972727272727, 0) { 0.000010 } ;
\node[circle, font=\tiny, draw=black, inner sep=2] (2) at (1.94545454545, 0.8) { 2 } ;
\draw[] (2) -- (l1) ;
\node[draw=black, font=\tiny, inner sep=2, fill=HighlightColor3] (l2) at (1.94545454545, 0) { 0.000011 } ;
\draw[] (2) -- (l2) ;
\node[draw=black, font=\tiny, inner sep=2, fill=HighlightColor4] (l3) at (2.91818181818, 0) { 0.0001 } ;
\coordinate (p1) at (3.40454545455, -1.5);
\draw[thick, black] (3.4045454545454543, -0.95) -- (3.40454545455, 2.85) ;
\node[circle, font=\tiny, draw=black, inner sep=2] (3) at (2.91818181818, 2.4) { 3 } ;
\draw[] (3) -- (l3) ;
\node[draw=black, font=\tiny, inner sep=2, fill=GoogleColor] (l4) at (3.89090909091, 0) { 0.000 } ;
\node[circle, font=\tiny, draw=black, inner sep=2] (5) at (4.86363636364, 0.8) { 5 } ;
\draw[] (5) -- (l4) ;
\node[draw=black, font=\tiny, inner sep=2, fill=HighlightColor5] (l5) at (4.86363636364, 0) { 0.001 } ;
\coordinate (p2) at (5.35, -1.5);
\draw[thick, black] (5.35, -0.95) -- (5.35, 2.85) ;
\draw[] (5) -- (l5) ;
\node[draw=black, font=\tiny, inner sep=2, fill=AmazonColor] (l6) at (5.83636363636, 0) { 0.0011 } ;
\coordinate (p3) at (6.32272727273, -1.5);
\draw[thick, black] (6.322727272727272, -0.95) -- (6.32272727273, 2.85) ;
\node[circle, font=\tiny, draw=black, inner sep=2] (6) at (5.83636363636, 2.4) { 6 } ;
\draw[] (6) -- (l6) ;
\node[draw=black, font=\tiny, inner sep=2, fill=AMDColor] (l7) at (6.80909090909, 0) { 0.0100 } ;
\node[circle, font=\tiny, draw=black, inner sep=2] (8) at (7.78181818182, 0.8) { 8 } ;
\draw[] (8) -- (l7) ;
\node[draw=black, font=\tiny, inner sep=2, fill=IntelColor] (l8) at (7.78181818182, 0) { 0.0101 } ;
\coordinate (p4) at (8.26818181818, -1.5);
\coordinate (p5) at (8.26818181818, -1.5);
\coordinate (p6) at (8.26818181818, -1.5);
\draw[thick, black] (8.268181818181818, -0.95) -- (8.26818181818, 2.85) ;
\draw[] (8) -- (l8) ;
\node[draw=black, font=\tiny, inner sep=2, fill=AppleColor] (l9) at (8.75454545455, 0) { 0.1 } ;
\coordinate (p7) at (9.24090909091, -1.5);
\draw[thick, black] (9.24090909090909, -0.95) -- (9.24090909091, 2.85) ;
\node[circle, font=\tiny, draw=black, inner sep=2] (9) at (8.75454545455, 2.4) { 9 } ;
\draw[] (9) -- (l9) ;
\node[draw=black, font=\tiny, inner sep=2, fill=HighlightColor] (l10) at (9.72727272727, 0) { 0.100101 } ;
\coordinate (p8) at (10.2136363636, -1.5);
\draw[thick, black] (10.213636363636363, -0.95) -- (10.2136363636, 2.85) ;
\node[circle, font=\tiny, draw=black, inner sep=2] (10) at (9.72727272727, 2.4) { 10 } ;
\draw[] (10) -- (l10) ;
\node[draw=black, font=\tiny, inner sep=2, fill=HighlightColor2] (l11) at (10.7, 0) { 0.101 } ;
\coordinate (p9) at (11.1863636364, -1.5);
\coordinate (p10) at (11.1863636364, -1.5);
\coordinate (p11) at (11.1863636364, -1.5);
\coordinate (p12) at (11.1863636364, -1.5);
\draw[thick, black] (11.186363636363636, -0.95) -- (11.1863636364, 2.85) ;
\node[circle, font=\tiny, draw=black, inner sep=2] (11) at (10.7, 2.4) { 11 } ;
\draw[] (11) -- (l11) ;
\draw[] (1) -- (2) ;
\node[circle, font=\tiny, draw=black, inner sep=2] (4) at (3.89090909091, 2.4) { 4 } ;
\draw[] (4) -- (5) ;
\node[circle, font=\tiny, draw=black, inner sep=2] (7) at (6.80909090909, 2.4) { 7 } ;
\draw[] (7) -- (8) ;
\draw[] (3) -- (1) ;
\node[circle, font=\tiny, draw=black, inner sep=2] (0) at (0.0, 2.4) { 0 } ;
\draw[] (0) -- (3) ;
\draw[red, opacity=0.0] (-0.2, -0.2) rectangle (10.9, 2.6) ;
\fill[HighlightColor] (-0.5, -1.7999999999999998) rectangle (0.0176991150442, -2.8) node[white, font=\tiny, pos=0.5] (b0) { 0 } ;
\fill[HighlightColor2] (0.017699115044247704, -1.7999999999999998) rectangle (0.121238938053, -2.8) node[white, font=\tiny, pos=0.5] (b1) { 1 } ;
\fill[HighlightColor3] (0.12123893805309727, -1.7999999999999998) rectangle (0.43185840708, -2.8) node[white, font=\tiny, pos=0.5] (b2) { 2 } ;
\fill[HighlightColor4] (0.43185840707964596, -1.7999999999999998) rectangle (0.638938053097, -2.8) node[white, font=\tiny, pos=0.5] (b3) { 3 } ;
\fill[GoogleColor] (0.638938053097345, -1.7999999999999998) rectangle (1.26017699115, -2.8) node[white, font=\tiny, pos=0.5] (b4) { 4 } ;
\fill[HighlightColor5] (1.2601769911504426, -1.7999999999999998) rectangle (1.8814159292, -2.8) node[white, font=\tiny, pos=0.5] (b5) { 5 } ;
\fill[AmazonColor] (1.88141592920354, -1.7999999999999998) rectangle (2.60619469027, -2.8) node[white, font=\tiny, pos=0.5] (b6) { 6 } ;
\fill[AMDColor] (2.606194690265487, -1.7999999999999998) rectangle (3.22743362832, -2.8) node[white, font=\tiny, pos=0.5] (b7) { 7 } ;
\fill[IntelColor] (3.227433628318584, -1.7999999999999998) rectangle (6.12654867257, -2.8) node[white, font=\tiny, pos=0.5] (b8) { 8 } ;
\fill[AppleColor] (6.126548672566371, -1.7999999999999998) rectangle (6.33362831858, -2.8) node[white, font=\tiny, pos=0.5] (b9) { 9 } ;
\fill[HighlightColor] (6.33362831858407, -1.7999999999999998) rectangle (7.78318584071, -2.8) node[white, font=\tiny, pos=0.5] (b10) { 10 } ;
\fill[HighlightColor2] (7.783185840707963, -1.7999999999999998) rectangle (11.2, -2.8) node[white, font=\tiny, pos=0.5] (b11) { 11 } ;
\draw[thick, fill=black, fill opacity=0.1] (-0.5, -1.6999999999999997) rectangle (11.2, -2.9) ;
\draw[thick, black] (-0.5, -1.6999999999999997) -- (-0.5, -2.9) ;
\draw[thick, black] (0.47499999999999987, -1.6999999999999997) -- (0.475, -2.9) ;
\draw[thick, black] (1.4499999999999997, -1.6999999999999997) -- (1.45, -2.9) ;
\draw[thick, black] (2.425, -1.6999999999999997) -- (2.425, -2.9) ;
\draw[thick, black] (3.3999999999999995, -1.6999999999999997) -- (3.4, -2.9) ;
\draw[thick, black] (4.375, -1.6999999999999997) -- (4.375, -2.9) ;
\draw[thick, black] (5.35, -1.6999999999999997) -- (5.35, -2.9) ;
\draw[thick, black] (6.325, -1.6999999999999997) -- (6.325, -2.9) ;
\draw[thick, black] (7.299999999999999, -1.6999999999999997) -- (7.3, -2.9) ;
\draw[thick, black] (8.274999999999999, -1.6999999999999997) -- (8.275, -2.9) ;
\draw[thick, black] (9.25, -1.6999999999999997) -- (9.25, -2.9) ;
\draw[thick, black] (10.225, -1.6999999999999997) -- (10.225, -2.9) ;
\end{tikzpicture}

	\caption{A radix tree forest built with \Cref{alg:radix_forest}. Note that all root nodes only have a right
	child. We manually set the reference for the left child to its left neighbor since it in practice almost always overlaps the left boundary. During
	sampling, the decision whether the left or right child must be used is purely based on the cumulative distribution function used
	as an input for the construction: Each node checks whether $\xi$ is smaller than the cumulative value with the same index.}
	\label{Fig:radix_forest}
\end{figure}
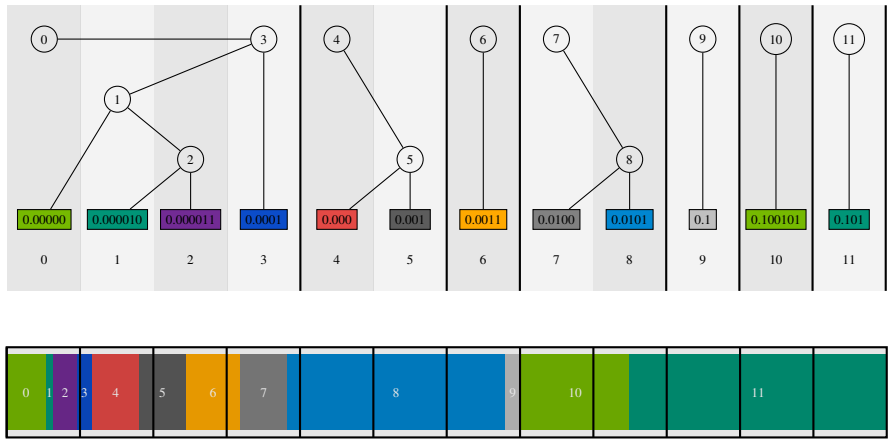

\begin{minicg}
	tree = RadixTree()
	N = 1000
	probs_construction = [((x*x+1) * random.random() * 0.05 if x < N / 2 else (x*x+1) * random.random()) for x in range(N)]
	cdf_construction = []
	cur = 0
	for p in probs_construction:
		cur += p
		cdf_construction.append(cur)
	for i in range(len(probs_construction)):
		probs_construction[i] /= cur
		cdf_construction[i] /= cur
	tree.build(probs_construction)
	tree.calc_leaf_levels()
	M = 32 #1024 * 1024
	avg = 0
	avg_balanced = 0
	avg32 = 0
	avg32_balanced = 0
	for i in range(M/32):
		m = 0
		m2 = 0
		for j in range(32):
			xi = random.random()
			c = tree.calc_num_steps(xi)
			c2 = tree.calc_num_steps_balanced(xi)
			m = max(m, c)
			m2 = max(m2, c2)
			avg += c
			avg_balanced += c2
		avg32 += m
		avg32_balanced += m2
	avg /= float(M)
	avg32 /= float(M/32)
	avg_balanced /= float(M)
	avg32_balanced /= float(M/32)

	# print '\n\n'
	# Tikz.begin()
	# tree.build(probs_construction, False, xscale, yscale, False, False, '', 0.0, True, colors)
	# Tikz.end()

	# print 'avg = {}\n'.format(avg)
	# print 'avg32 = {}\n'.format(avg32)
	# print 'avg balanced = {}\n'.format(avg_balanced)
	# print 'avg32 balanced = {}\n'.format(avg32_balanced)
\end{minicg}
%
% show construction = False

%

We indicate that a cell in the guide table is only overlapped by a single interval by setting the reference stored in
the cell to the two's complement of the index of the interval. Then, a most significant bit set to one identifies such an interval, and one
unnecessary indirection is avoided. If the size of the distribution is
sufficiently small, further information could also be stored in the reference, such as a flag that there are exactly two
intervals that overlap the cell. Then, only one comparison must be performed and there is no need to explicitly store a
node. 

Building the radix forest is slightly faster than building a radix tree over the entire distribution since merging stops
earlier. However, we found the savings to be almost negligible, similarly to the effort required to set the references in
the guide table.

\Cref{alg:final_algo} shows the resulting combined method that maps $\xi$ to $i$. First, it looks up a reference in the guide table at the index $g = \lfloor M
\cdot \xi \rfloor$. If the most significant bit of the reference is one, $i$ is the two's complement of this reference. Otherwise
the reference is the root node of the tree which is traversed by iteratively comparing $\xi$ to the value of the
cumulative distribution function with the same index as the current node, advancing to its left child if $\xi$ is smaller, and to
its right child if $\xi$ is larger or equal. Tree traversal again terminates if the most significant bit of the next node index
is one. As before, $i$ is determined by taking the two's complement of the index.

\begin{algorithm}
	\begin{algorithmic}
		\STATE \textbf{Input:} $\xi \in \left[0, 1\right)$, $data \in \left[0, 1\right)^n$ in increasing order, number of partitions $m$, $nodes$ created
		with \Cref{alg:radix_forest}, guide table $table$.
		\STATE \textbf{Output:} $i \in \{0, 1, ..., N - 1\}$.
		\STATE $g \gets \lfloor \xi \cdot m \rfloor$
		\STATE $j \gets table[g]$
		\WHILE{$msb(j) \neq 1$}
			\IF{$\xi < data[j]$}
				\STATE $j \gets nodes[j].child[0]$
			\ELSE
				\STATE $j \gets nodes[j].child[1]$
			\ENDIF
		\ENDWHILE
		\RETURN $\sim j$ 
	\end{algorithmic}
	\caption{Mapping $\xi$ to $i$ with the presented method combining a guide table with a radix forest.}
	\label{alg:final_algo}
\end{algorithm}

\section{Results} \label{Sec:Results}

We evaluate our sampling method in two steps: First, we compare to an Alias Method in order to quantify the impact on
convergence speed. Second, we compare to the Cutpoint Method with binary search which performs the identical
mapping and therefore allows one to quantify execution speed.

\Cref{Fig:2d_zoom} illustrates how sampling is affected by the discontinuities of the Alias Method.
The results indicate that sampling with the Alias Method may indeed be less efficient.

\begin{figure}[h]
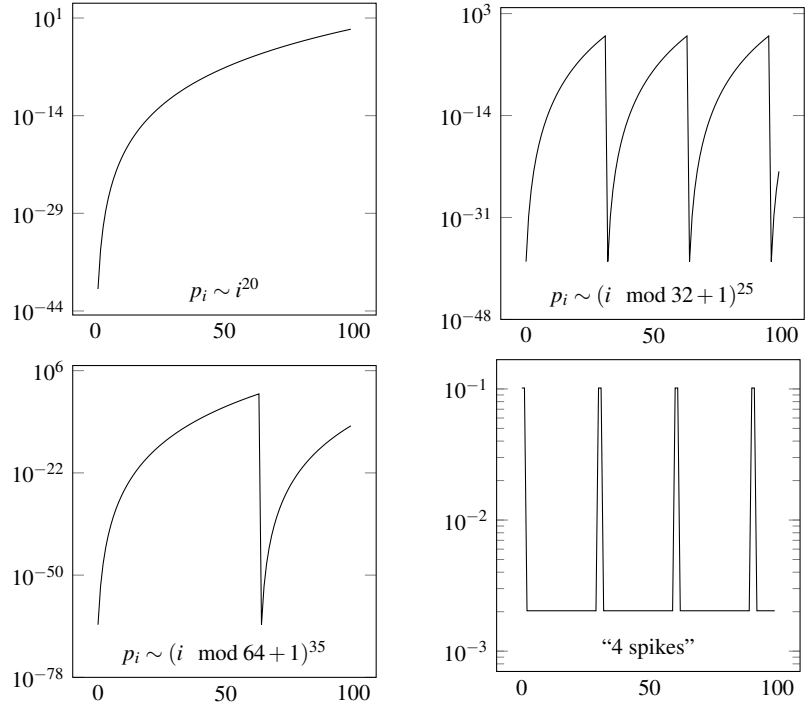

	\begin{subfigure}{0.48\textwidth}
		\centering
		\begin{tikzpicture}
			\begin{axis}[
				xtick style={draw=none},
				ymode=log,
				legend style={at={(0.5,0.01)},anchor=south, fill=none, font=\small, legend style={draw=none}},
				legend cell align=left,
				width=\textwidth,
				height=.3\textheight]
				\addlegendimage{empty legend}
				\addlegendentry{$p_i \sim i^{20}$}
				\input{pdf1.txt}
			\end{axis}
		\end{tikzpicture}
	\end{subfigure}
	\begin{subfigure}{0.48\textwidth}
		\centering
		\begin{tikzpicture}
			\begin{axis}[
				xtick style={draw=none},
				ymin = 10e-49,
				ymode=log,
				legend style={at={(0.5,0.01)},anchor=south, fill=none, font=\small, legend style={draw=none}},
				legend cell align=left,
				width=\textwidth,
				height=.3\textheight]
				\addlegendimage{empty legend}
				\addlegendentry{$p_i \sim (i\mod 32 +1)^{25}$}
				\input{pdf2.txt}
			\end{axis}
		\end{tikzpicture}
	\end{subfigure}\hfill\\
	\begin{subfigure}{0.48\textwidth}
		\centering
		\begin{tikzpicture}
			\begin{axis}[
				xtick style={draw=none},
				ymode=log,
				ymin = 10e-79,
				legend style={at={(0.5,0.01)},anchor=south, fill=none, font=\small, legend style={draw=none}},
				legend cell align=left,
				width=\textwidth,
				height=.3\textheight]
				\addlegendimage{empty legend}
				\addlegendentry{$p_i \sim (i\mod 64 +1)^{35}$}
				\input{pdf3.txt}
			\end{axis}
		\end{tikzpicture}
	\end{subfigure}
	\begin{subfigure}{0.48\textwidth}
		\centering
		\begin{tikzpicture}
			\begin{axis}[
				xtick style={draw=none},
				ymode=log,
				ymin = 0.0007, %
				legend style={at={(0.5,0.01)},anchor=south, fill=none, font=\small, legend style={draw=none}},
				legend cell align=left,
				width=\textwidth,
				height=.3\textheight]
				\addlegendimage{empty legend}
				\addlegendentry{``4 spikes''}
				\input{pdf4.txt}
			\end{axis}
		\end{tikzpicture}
	\end{subfigure}
	\caption{Example distributions for the numerical results in \Cref{Tab:Evaluation}.}
	\label{fig:evaluation_pdf1}
\end{figure}

\begin{table}
	\caption{Measuring the maximum and average number of memory load operations required for searching as well as the
	average number of load operations or idle operations of 32 simulations that need to be synchronized (average$_{32}$) shows that
	while the maximum number of load operations is increased, for distributions with a high range the average number of
	load operations is typically reduced.}
	\label{Tab:Evaluation}
	\begin{tabular}{l@{\hskip .95in}r@{\hskip .4in}r@{\hskip .4in}r}
		$p_i \sim i^{20}$ & maximum & average & average$_{32}$\\
		\hline
		Cutpoint Method + binary search & 8 & 1.25 & 3.66\\
		Cutpoint Method + radix forest & 16 & 1.23 & 3.46\\\\
		$p_i \sim (i\mod 32 +1)^{25}$ & maximum & average & average$_{32}$\\
		\hline
		Cutpoint Method + binary search & 6 & 1.30 & 4.62\\
		Cutpoint Method + radix forest & 13 & 1.22 & 3.72\\\\
		$p_i \sim (i\mod 64 +1)^{35}$& maximum & average & average$_{32}$\\
		\hline
		Cutpoint Method + binary search & 7 & 1.19 & 4.33\\
		Cutpoint Method + radix forest  & 13 & 1.11 & 2.46\\\\
		``4 spikes'' & maximum & average & average$_{32}$\\ %
		\hline
		Cutpoint Method + binary search & 4 & 1.60 & 3.98\\
		Cutpoint Method + radix forest & 5 & 1.67 & 4.93
	\end{tabular}
\end{table}

\Cref{Tab:Evaluation} details the performance improvement of our method as compared to the Cutpoint Method with binary
search for the distributions shown in \Cref{fig:evaluation_pdf1}. As expected, the sampling performance of the new method
is similar to the Cutpoint Method with binary search, which shares the same primary acceleration data
structure, the guide table. For reasonable table sizes both perform almost as good as sampling with an Alias Method,
however, do so without affecting the distribution quality.

Sampling densities with a high dynamic range can be efficiently accelerated using the Cutpoint Method and its guide
table. However, since then some of its cells contain many small values with largely different magnitudes, performance suffers from
efficiency issues of index bisection. Radix tree forests improve on this aspect by storing an explicit tree (see the
parallel \Cref{alg:radix_forest}).

On a single processor with strictly serial execution, our method marginally improves the average search time as compared to the Cutpoint Method with binary search since the
overall time is largely dominated by the time required to find large values. For every value that can be directly
determined from the guide table -- since it is the only one in a cell -- this process is already optimal. In some cases
the average search time of our new method can be slightly slower since manually assigning the index of interval overlapping %
from the left as the left child of the root node the tree deteriorates overall tree quality (see \Cref{Tab:Evaluation}, example ``4 spikes''). On the other
hand, performing bisection to find a value inside a cell that includes many other values does not take the underlying
distribution into account and is therefore suboptimal. Still, since these values are sampled with a low probability, the
impact on the average sampling time is low.

The optimizations for values with a lower probability become more important in parallel
simulations where the slowest simulation determines the run time for every group. The ``4 spikes'' example is a synthetic
bad case for our method: Spikes are efficiently sampled only using the guide table, whereas all other values have
uniform probabilities. Therefore binary search is optimal. The explicit tree, on the other hand, always has one
sub-optimal first split to account for intervals overlapping from the left, and therefore requires one additional
operation.

In practice, parallel execution of the sampling process often requires synchronization, and therefore suffers more from
the outliers: Then, the slowest sampling process determines the speed of the entire group that is synchronized. On
Graphics Processing Units (GPUs), such groups are typically of size 32. Under these circumstances our method performs
significantly better for distributions with a high dynamic range (see \Cref{Tab:Evaluation}, third column).

It is important to note that if the maximum execution time is of concern, binary search almost always achieves the best worst case
performance. Then explicit tree structures can only improve the average time if the maximum depth does not exceed the
number of comparisons required for binary search, which is the binary logarithm of the number of elements.

\section{Discussion} \label{Sec:Discussion}

A multi-dimensional inversion method proceeds component by component; the two-dimensional distributions in
\Cref{Fig:2d_zoom} have been sampled by first calculating the cumulative distribution function of the image rows and one cumulative
distribution function for each row. After selecting a row its cumulative distribution function is sampled to select the column.
Finally, as distributions are considered piecewise constant, a sub-pixel position, i.e. a position in the constant piece, is required.
Therefore, the relative position in the pixel is calculated by rescaling the relative position in the row/column to the
unit interval. Other approximations, such as piecewise linear or piecewise quadratic require an additional, simple
transformation of the relative position \cite{AliasExtended}.

Building multiple tables and trees simultaneously, e.g. for two-dimensional distributions, is as simple as adding yet
another criterion to the extended check in \Cref{alg:radix_forest}: If the index of the left or right neighbor goes beyond the \emph{index
boundary} of a row, it is a leftmost or a rightmost node, respectively.

\Cref{alg:radix_forest} constructs binary trees. Due to memory access granularity, it may be beneficial to
construct 4-ary or even wider trees. A higher branching factor simply
results by just collapsing two (or more) levels of the binary trees. 

For reasonable table sizes, the Cutpoint Method with binary search
preserves the properties of the input samples at a
memory footprint comparable to the one required for the Alias Method.
Radix forest trees require additional memory to reference two children for each
value $p_i$ of the distribution.

Depending on the application and space constraints, it may be beneficial to use balanced trees instead of radix trees.
Balanced trees do not need to be built; their structure is implicitly defined, and for each cell we only need to
determine the first and last interval that overlap it. Then, the implicit balanced tree is traversed by consecutive
bisection of the index interval.

\section{Conclusion}

Radix tree forests trade additional memory for a faster average case search
and come with a massively parallel construction algorithm with optimal load balancing
independent of the probability density function $p$.

While the performance of evaluating the inverse cumulative distribution function $P^{-1}$
is improved for highly nonuniform distributions $p$ with a
high dynamic range, performance is slightly worse
on distributions which can already be efficiently sampled using the Cutpoint
Method with binary search. %

Thus the choice of the best sampling algorithm depends
on the actual application. Our use case of illuminating computer generated scenes
by high dynamic range video environment maps greatly benefits from the
massively parallel construction and sampling of radix tree forests.

In future work, we will investigate the use of the massively parallel
algorithms for subsampling the activations of neural networks in order to
increase the efficiency of both training and inference.

\section*{Acknowledgements}

The authors would like to thank Carsten W\"achter and Matthias Raab for the discussion of the issues
of the Alias Method when used with low discrepancy sequences that lead to the development of radix tree forests.

\end{document}